# Main Manuscript for

# Revealing buried ferroelectric topologies by depth-resolved electron diffraction imaging.

Ting-Ran Liu[1], Koushik Jagadish[1], Xiangwei Guo[2,3], Maya Ramesh[4], Peter Meisenheimer[5], Harish Kumarasubramanian[1], Sajid Husain[5], Ann V. Ngo[1], Amir Avishai[6], Jayakanth Ravichandran[1,6,7], Darrell G. Schlom[4], Ramamoorthy Ramesh[5], Yu-Tsun Shao[1,6]*

[1]Mork Family Department of Chemical Engineering and Materials Science, University of Southern California, Los Angeles, CA 90089, USA.
[2]Department of Materials Science and Engineering, University of Wisconsin-Madison, Madison, WI 53706, USA.
[3]Department of Mechanical Engineering, University of Michigan, Dearborn, MI 48128, USA.
[4]Department of Materials Science and Engineering, Cornell University, Ithaca, NY 14850, USA.
[5]Department of Materials Science and Engineering, University of California, Berkeley, CA 94720, USA.
[6]Core Center of Excellence in Nano Imaging, University of Southern California, Los Angeles, CA 90089, USA.
[7]Ming Hsieh Department of Electrical and Computer Engineering, University of Southern California, Los Angeles, CA 90089, USA.

*To whom correspondence may be addressed: yutsunsh@usc.edu

**Author Contributions:** Y.-T.S. designed research; Y.-T.S., T.-R.L. K.J., X.G., M.R., H.K., S.H., and A.V.N. performed research; Y.-T.S., T.-R. L., K.J., X.G., M.R., P.M., H.K., S.H., and A.A. contributed new reagents/analytic tools; Y.-T.S., T.-R.L., X.G., A.V.N., J.R., D.G.S., and R.R. analyzed data; Y.-T.S. and T.-R.L. wrote the paper.

**Competing Interest Statement:** The authors declare no conflict of interest.

**This Word file includes:**

Main Text

Figures 1 to 4

## Abstract

Nanoscale topological polar textures promise new functionalities for ferroelectric memories and logic[1], yet their three-dimensional structure and mesoscale organization remain experimentally inaccessible. Here we introduce depth-resolved electron diffraction imaging (DREDI), a fast, non-destructive, method that maps polarization with <50 nm lateral and <10 nm depth sensitivity within fraction of a second. Its high acquisition speed enables the first continuous polarization mapping across six orders of magnitude in length scale, from nanometers to millimeters. Using epitaxial $BiFeO_3$ films, DREDI reveals a hidden depth evolution of polar textures: surface 71° stripes evolve into subsurface flux-closure vortices that bifurcate into three-fold vertices near the bottom interface. Cross-sectional multi-slice electron ptychography and phase-field modeling confirm these buried configurations and attribute them to strain heterogeneity and ferroelastic twinning in the $SrRuO_3$ electrode. Large-area analysis further shows that vertex-like frustration forms a mesoscale percolating network above a critical length scale of 4 µm. DREDI enables real-time, volumetric studies of buried topological textures in ferroic nanomaterials.

## Introduction

Topological defects are ubiquitous in ordered matter: domain walls, vortices, and skyrmions define how an order parameter field twists through space[2,3]. In ferroelectrics, such textures emerge from the intricate interplay of elastic, electrostatic, and gradient energies, giving rise to flux-closure domains, vortices[4-6], skyrmions[7,8]. These polar textures exhibit emergent properties absent in uniform single-domain states, including chirality[9,10] and negative capacitance[11,12]. Critically, these nanoscale textures are inherently three-dimensional (3D): their topology, stability, and switching behavior depend on how polarization evolves through the film thickness, where mechanical and electrostatic boundary conditions compete. Yet nearly all experimental insight to date has come from two-dimensional (2D) measurements or imaging, leaving the intrinsic depth evolution of polar textures largely unexplored.

Equally important for device-relevant ferroelectric films, the mesoscale arrangement of nanoscale textures is often as important as the textures themselves. Switching behavior, fatigue, and device-to-device variability all depend on how local domains connect and evolve across lateral distances. However, no existing metrology technique can rapidly map ferroic domain patterns across wafer-scale areas while retaining nanometer resolution, a capability urgently needed for emerging ferroelectric memory and logic technologies.

Most existing probes face fundamental limitations. Surface-sensitive techniques such as piezoresponse force microscopy (PFM), including vector PFM[4] and quadrature phase differential interferometry (QPDI)[13], have become powerful and widely used tools for imaging ferroelectric domain structures with high spatial resolution at sample surfaces. However, because the signal arises from local electromechanical interactions between the probe tip and the sample surface, PFM primarily probes near-surface polarization and can be less sensitive to buried or subsurface domain structures. Techniques capable of providing depth-sensitive polarization information over larger length scales are therefore highly desirable. On the other hand, aberration-corrected scanning transmission electron microscopy (S/TEM) enables direct imaging of polarization through atomic-scale displacements or diffraction contrast projected on a 2D plane. Conventional 3D methods, including tomographic AFM[14], serial focused ion beam and scanning electron microscopy (FIB-SEM)[15], and atom probe tomography (APT)[16], require physical sectioning or thinning, which often relaxes mechanical clamping and alters electrostatic screening. Such modifications can reconfigure the mesoscale structure of polarization textures themselves, for example creating polar vortices from trivial domains during cross-section TEM lamellae preparation[17]. Thus, a non-destructive volumetric imaging technique that preserves native boundary conditions and resolves 3D dipolar textures at the nanoscale remains an unmet need.

To address this challenge, we chose a model system of epitaxial $BiFeO_3$ (BFO) thin films, a multiferroic with rich domains physics and sensitivity to boundary conditions. The room-temperature multiferroic BFO exhibits coupling between polarization, strain, magnetism, and oxygen octahedral rotations, which led to diverse near-ground-state phases with tunable electronic properties[18–20]. In its rhombohedral ground state (*R3c*, space group No. 161), BFO exhibit spontaneous polarization along eight pseudocubic (pc) $<111>_{pc}$ directions, which can have 71°, 109°, or 180° domain walls[21,22]. Depending on strain state and electrostatic boundary conditions, epitaxial BFO films can exhibit periodic stripes, flux-closure domains, and even vortex-antivortex arrays[23–25]. Prior studies, however, have largely inferred such structures from surface probes or cross-sectioned lamellae, leaving their intrinsic depth evolution unresolved[25–29]. This rich tunability makes BFO an ideal platform for exploring how complex polarization textures evolve beneath the surface and testbed for validating a new non-destructive, volumetric imaging approach.

Here, we establish depth-resolved electron diffraction imaging (DREDI) as a non-destructive SEM-based approach for probing buried polarization structures with nanometer-scale lateral and <10 nm depth sensitivity within fraction of a second, which is 1,000× faster acquisition than scanning-probe approaches. DREDI leverages dynamical diffraction effects encoded in Kikuchi band intensity asymmetries to determine local polarization orientation, while systematic tuning of the electron beam energy provides tomographic sensitivity. Combined with automated tile-and-stitch workflow, we demonstrate, for the first time, a continuous polarization mapping spanning six orders of magnitudes in length scale, from nanometers to millimeters.

Using this capability, we reveal the hidden depth evolution of polar textures in 30 nm $BiFeO_3$ films: from bifurcated three-fold domain-wall vertices near the bottom interface, transitioning to quadrant vortex domains in the film interior, which ultimately evolve into regular 71° stripe domains at the surface. Phase-field modeling attributes these buried structures to strain heterogeneities and ferroelastic twinning in the $SrRuO_3$ bottom electrode, while cross-sectional multislice electron ptychography independently confirms the presence of such complex 3D domain structures. Finally, large-area DREDI maps reveal that these frustrated, vertex-like regions are not isolated nanoscale anomalies but instead form a mesoscale percolating network extending over tens to hundreds of micrometers. This establishes DREDI as a powerful platform for connecting nanoscale ferroic textures with their mesoscale and macroscale organization.

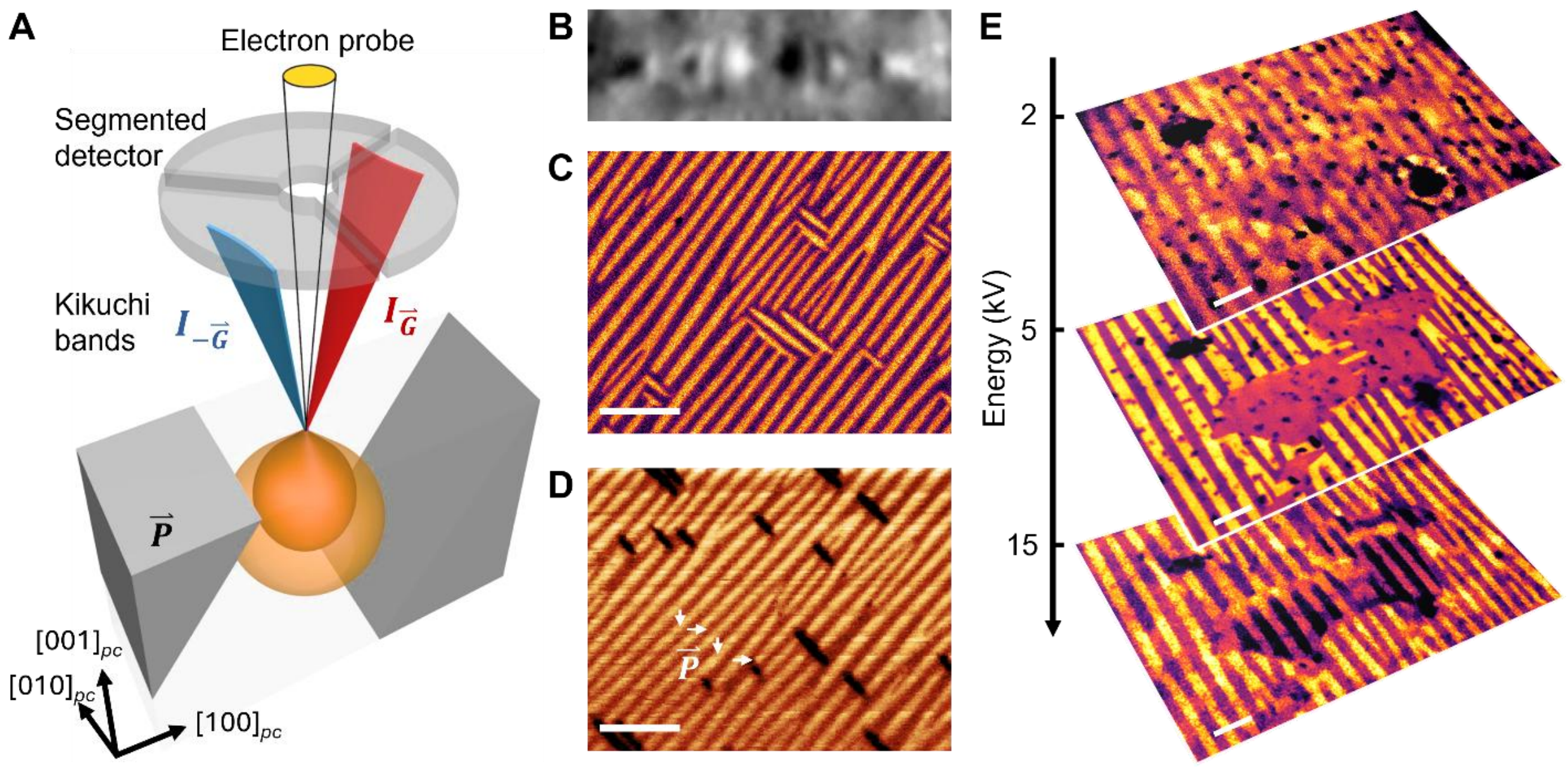

**Fig. 1| DREDI of embedded 3D polarization domains in a $BiFeO_3$ thin film. (A)** Schematic illustration of the DREDI setup implemented in an SEM. Intensity asymmetry of polarity-sensitive Kikuchi bands can be captured using a segmented DBS detector. **(B)** Full momentum-space Kikuchi pattern, where the (104) band exhibits a strong intensity asymmetry. The difference between Kikuchi patterns from two neighboring ferroelectric domains is shown. **(C,D)** Polarization maps obtained by DREDI (C) and PFM (D) showing 71˚ domains. **(E)** Depth-resolved DREDI maps acquired at different landing energies of 2 kV, 5 kV, and 15 kV, revealing embedded 3D polarization domain patterns resembling a capybara. Scale bars: 1 µm (C,D); 500 nm (E). The arrows indicate polarization directions, and 'pc' denotes pseudocubic.

## Results

### Principle and experimental setup of DREDI

**Figure 1A** shows the schematic experimental implementation of DREDI in the widely accessible SEM, performed on as-grown 30 nm $BiFeO_3$/30 nm $SrRuO_3$/$DyScO_3$ thin film heterostructure without surface treatment nor conductive coating. The principle of DREDI is based on two aspects of multiple electron scattering: (1) Friedel symmetry breaking in Kikuchi bands due to dynamical diffraction effects, and (2) multiple inelastic scattering events that define the range of electrons within the sample. Kikuchi bands form in SEM when thermal diffuse scattered electrons satisfy the Bragg diffraction condition, producing characteristic band patterns that reflect the underlying crystal symmetry. In a centrosymmetric lattice, the Friedel pairs of Kikuchi bands ($\pm\vec{G}$) exhibit equal intensities. However, in non-centrosymmetric crystals like $BiFeO_3$, the charge redistribution associated with ferroelectric polarization leads to breakdown of Friedel's law owing to dynamical diffraction effects[30–32].

To substantiate this mechanism, we performed Bloch-wave dynamical diffraction simulations of electron channeling patterns (ECP) for $BiFeO_3$ using EMsoft[33]. The simulations, carried out under the same experimental geometry (zero sample tilt along the $[001]_{pc}$ growth direction), identify specific kikuchi bands associated with ferroelectric polarization **(Fig. S1, Supplementary Note 1.**

**Polarization Direction)**. Accordingly, the polarization direction can be determined from the intensity differences of these polarity-sensitive Kikuchi bands.

Experimentally, this can be recorded by electron backscattered diffraction[34,35] (EBSD, **Figs. 1B & S2**) or a segmented, directional backscatter detector (DBS**, Fig. S3**) akin to those used in differential phase contrast (DPC) imaging[36]. While EBSD captures the full momentum distribution of scattered electrons at each probe position, acquisition typically requires tens of minutes to a few hours. In contrast, in its DBS implementation, DREDI enables real-time readout of intensity asymmetry in Kikuchi bands within fractions of a second. **Figures 1C & 1D** show the comparison between PFM and DREDI imaging over comparable fields of view.

## Depth-resolved imaging of buried polarization nanodomains

An important feature unique to DREDI, as implemented in a SEM, is its ability to resolve polarization structures along the depth dimension non-destructively. By varying the incident electron energy, we systematically control the interaction volume of backscattered electrons. Low acceleration voltages (1-3 kV) are surface-sensitive, whereas higher energies (5-15 kV) progressively probe deeper layers of a 30 nm-thick $BiFeO_3$ film. **Figure 1E** shows DREDI images acquired at landing energies of 2 kV, 5 kV, and 15 kV, each revealing distinct domain patterns.

To quantify the depth dependence, we performed Monte Carlo simulations of backscattered electron trajectories in the 30 nm $BiFeO_3$/30 nm $SrRuO_3$/$DyScO_3$ heterostructure using CASINO[37]. The simulations show that the backscattered electron distribution shifts to greater depths and broadens with increasing landing energy (**Figs. S4 & S5, Supplementary Note 3. Depth Sensitivity**), with characteristic penetration depths (defined as the depth containing 90% of the landing energy) of approximately 4 nm at 2 kV, 20 nm at 5 kV, 32 nm at 10 kV, and 160 nm at 15 kV.

Left column of **Figure 2** (**Figs. 2A**, **2D**, **2G**) summarizes the experimentally observed depth-resolved DREDI measurements. Near the surface, the film exhibits regular 71° stripe domains. With increasing probe depth, these stripes evolve into quadrant vortex-like configurations, and finally bifurcate into a pair of three-fold vertices at higher voltage, revealing the presence of sub-surface, frustrated polarization textures. To our best knowledge, these results provide the first direct experimental evidence of embedded 3D topological polar textures within a multiferroic thin film.

## Phase-field simulations of domain evolution

To evaluate whether the experimentally observed domain configurations naturally emerge from standard ferroelectric energetics, we performed phase-field simulations to capture the mesoscale evolution of ferroelectric domains under realistic mechanical and electrostatic boundary conditions. The simulations are based on the time-dependent Ginzburg–Landau formalism and include contributions from elastic, electrostatic, and gradient energies[38,39]. The simulated geometry replicates the experimental heterostructure, consisting of a 30 nm $BiFeO_3$ layer grown on $SrRuO_3$-coated $DyScO_3$ **(Fig. S6A)**. Random initial polarization fluctuations were introduced to nucleate domains, and the system was allowed to evolve toward equilibrium (details see **Methods**).

Middle column of **Figure 2** shows the simulations reproducing the depth progression observed experimentally. Near the top surface (z = 1 nm), stripe-like 71° domains with alternating in-plane polarization components are stabilized (**Fig. 2B**), consistent with low-voltage DREDI. At intermediate depths (z = 19 nm), these stripes reorganize into a four-fold flux-closure vortex (**Fig. 2E**). This configuration consists entirely of 71° domain walls, remains stable due to their low formation energy and coherent strain compatibility. Near the bottom interface (z = 30 nm), where mechanical clamping and electrostatic screening dominate, the four-fold quadrants become unstable and bifurcates into two asymmetric three-fold vertices connected by a vertical 109° domain wall (**Fig. 2H**). The corresponding in-plane polarization curl maps $(\nabla \times P)_z$ (**Figs. 2C**, **2F**, **2I**) highlight the continuous topological transformation along the film thickness. The 3D rendering of domains walls shown in **Fig. 2J** confirms excellent agreement with depth-resolved DREDI, revealing a sub-surface, frustrated polarization domain network that transitions from stripes to vortex quadrants to vertex pairs.

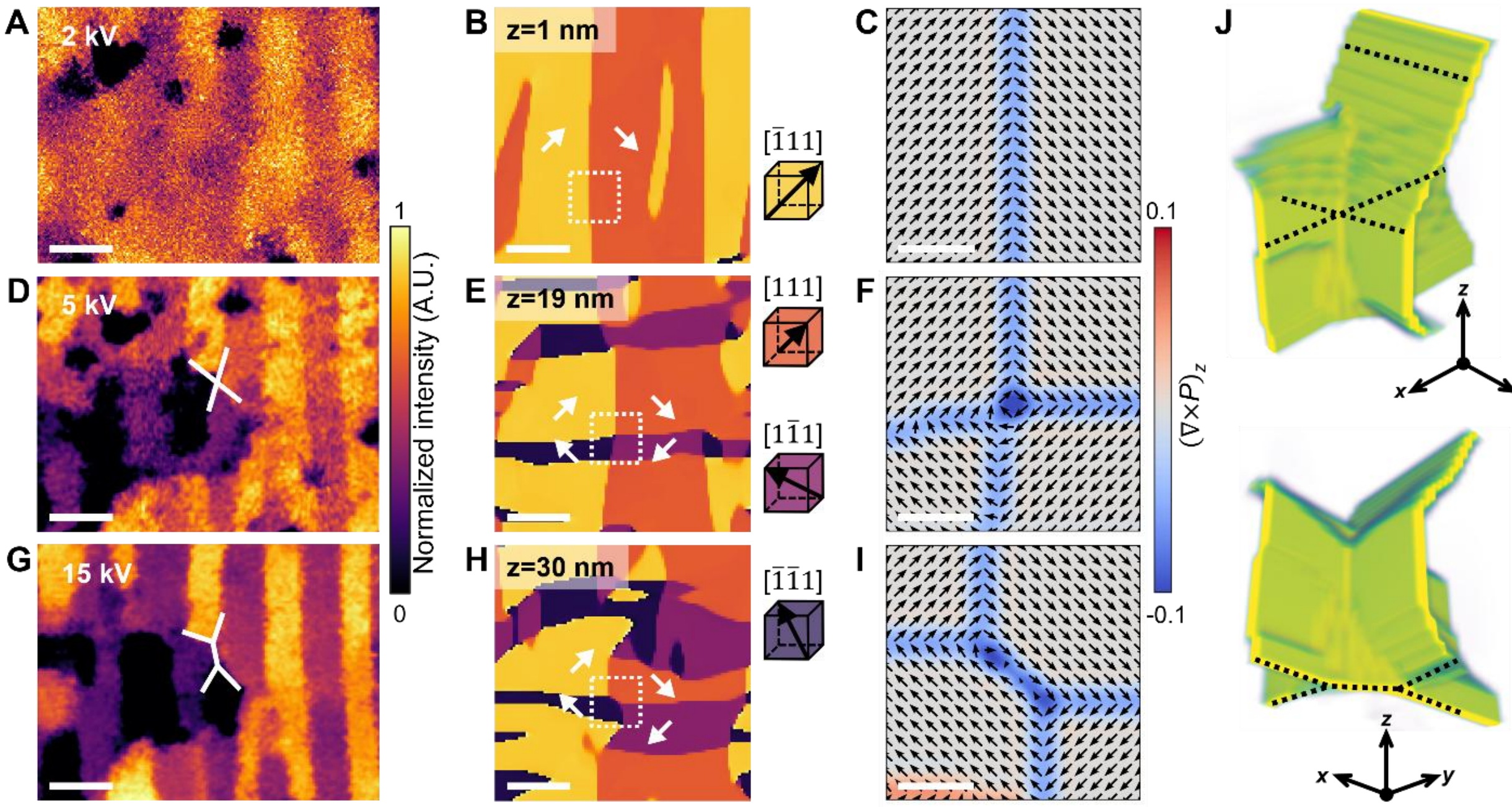

**Fig. 2| Depth evolution of stripe domains to vortex and vertices. (A, D, G)** DREDI images acquired at electron energies of 2kV (A), 5kV (D), and 15kV (G), showing embedded frustrated domain patterns beneath the surface. The white lines highlight boundaries between three- and four-fold domains. **(B, E, H)** Corresponding phase-field simulation of ferroelectric domain configurations at three representative planes along the film thickness (z = 1, 19, to 30 nm, from top to bottom). The white vectors show the in-plane polarization directions. The other colors represent different ferroelectric domains (corresponding polar directions as illustrated on the right). The simulation reveals a clear domain evolution from stripe-like states near the top (B), to a four-fold vortex configuration in the middle region (E), and finally to a pair of three-fold vertices near the bottom of the film (H). **(C, F, I)** Enlarged local polarization maps extracted from the phase-field simulations, corresponding to the boxed regions in (B, E, H), with color contrast representing the in-plane polarization curl $(\nabla \times P)_z$, highlighting the topological transformation associated with domains bifurcating along the film thickness. **(J)** 3D visualization of the domain wall network throughout the BFO film, showing stripe-like domains near the top and a frustrated region toward the bottom, where domain-wall splitting is observed. Scale bars: 1 μm (A, D, G), 50 nm (B, E, H), 10 nm (C, F, I).

## Validation with cross-sectional multislice electron ptychography

To independently verify the buried 3D polar textures revealed by depth-resolved DREDI, we performed multislice electron ptychography (MEP) on cross-sectional lamellae extracted from regions exhibiting frustrated subsurface domains in plan-view. MEP serves as a quantitative, atomic-resolution benchmark because it reconstructs the electrostatic potential by solving the inverse electron-scattering problem, providing quantitative access to both cation and anion positions with <0.2 Å lateral and ~2-3 nm depth resolution[40–42]. This makes MEP a powerful method for validating whether the complex 3D textures observed by DREDI correspond to intrinsic polarization structures within the film.

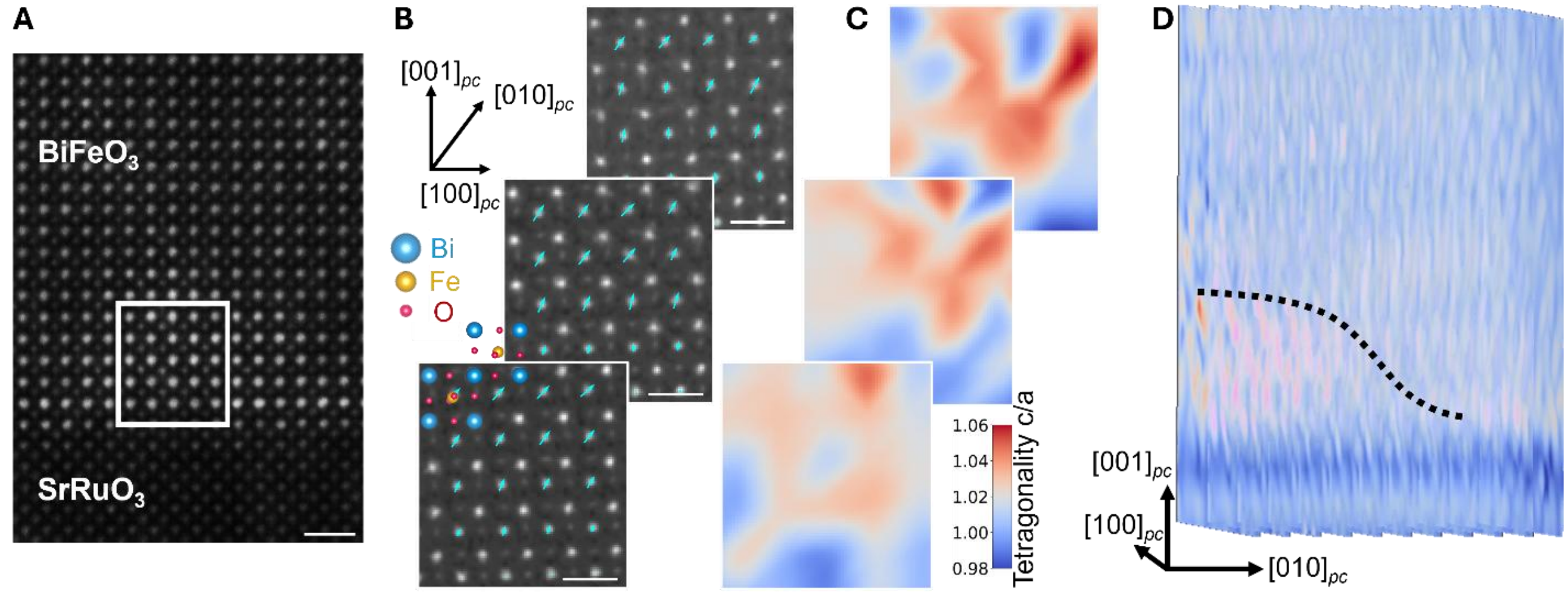

**Fig. 3| Cross-sectional STEM and MEP validation of vertex-like polar textures. (A)** LAADF-STEM image of the $BiFeO_3/SruRuO_3$ bottom interface along $[010]_{pc}$ showing a triangular region with bright contrast enclosed by two ~45˚ boundaries, obtained at similar frustrated regions identified by DREDI in plan-view. **(B,C)** Plane-by-plane MEP reconstruction of cations and anions in the boxed region of (A), showing depth-dependent polarization rotation (B) and lattice distortions (C), consistent with vertex-like configurations. Inset shows the corresponding $BiFeO_3$ structure model with *R3c* symmetry. Projected polarization $P_{XZ}$ derived from cation-anion relative displacements, while local tetragonality (c/a ratio) extracted from Bi-Bi distances. Here, c and a denote the out-of-plane and in-plane lattice parameters, respectively. **(D)** 3D rendering of the tetragonality field, showing strong spatial inhomogeneity indicated by black dashed curve, consistent with bifurcated vertex configurations observed by DREDI and phase-field modeling. Scale bars: 1 nm (A); 5Å (B).

**Figure 3** shows the polarization ($\vec{P}$) and lattice distortion maps obtained from cross-sectional MEP. Low-angle annular dark-field (LAADF) scanning transmission electron microscopy (STEM) reveals triangular regions of bright contrasts near the bottom $BiFeO_3/SrRuO_3$ interface, enclosed by two ~45˚ boundaries (**Fig. 3A**). These triangular regions resemble the domain bifurcation detected by DREDI at intermediate and high landing energies (Figs. 2G & 2J). Although this geometry nominally resembles the intersection of two symmetry-permissible 71˚ domain walls in the *R3c* structure[21,27], the internal structure differs substantially.

Depth-resolved MEP reconstructions show that the triangular region does not consist of simple 71˚ domain walls. Instead, the projected polarization vector $\boldsymbol{P_{XZ}}$ rotates continuously through the film thickness, and neighboring planes display significant variations in polar magnitude and direction (**Fig. 3B**). The measured displacements (~0.32±0.07 Å, corresponding to ~98 μC/cm$^2$ along one of the <111>$_{pc}$ using the Born effective charge model[28,43]) include planes where $\boldsymbol{P_{XZ}}$ rotates ~45° toward $[001]_{pc}$, (top panel, Fig. 3B), consistent with the vertex bifurcation observed by DREDI.

In addition to polarization, unit-cell tetragonality (*c/a*; *c* and *a* denote out-of-plane and in-plane lattice parameters, respectively) can be obtained from interatomic Bi-Bi distances in each plane. The resulting unit-cell tetragonality maps (**Fig. 3C**) and their 3D rendering (**Fig. 3D**) reveal inhomogeneous lattice distortions. Such lattice deformation is consistent with the lattice disclination field expected at the core of a vortex-vertex transition[44]. These signatures, together with the depth-dependent polarization rotation, confirm that the vortex-vertex domain configurations revealed by DREDI (**Fig. 2**) are real, intrinsic structural features of the $BiFeO_3$ film.

**Mapping polar order across multiple length scales**

After establishing that the vortex–vertex configurations identified by DREDI are intrinsic structural features, a natural next question emerges: how frequently do these frustrated regions occur, and how are they distributed across the film? Depth-resolved DREDI and cross-sectional MEP probe only local areas, yet many ferroic properties—such as switching behavior or device variability—are governed by domain organization across tens to hundreds of micrometers. Addressing this requires polarization mapping over areas many orders of magnitude larger than accessible by atomic-resolution STEM or PFM, while still retaining nanometer spatial resolution.

DREDI is uniquely suited for this task because it combines rapid acquisition with non-destructive imaging. Inspired by multiscale workflows in life-science electron microscopy[45], we implemented an automated tile-and-stitch acquisition protocol to obtain continuous, nanometer-resolution polarization maps over macroscopic fields of view (see **Methods**). As shown in **Figure 4A**, this approach enables continuous mapping of the in-plane polarization-orientation map spanning over 100 μm, while preserving nanoscale structural detail (and up to 1 mm field of view in **Supplementary Video 1**). The fast Fourier transform (FFT) of the full field confirms an average stripe domain periodicity of 183 nm (**Fig. 4A**, bottom-right inset), consistent with anisotropic in-plane epitaxial strain imposed by the $DyScO_3$ substrate[46,47].

Upon inspecting the data, however, we noticed spurious mesoscale pockets of polar frustration, where vertical and horizontal 71° stripe domains coexist, analogous to the frustrated subsurface textures uncovered by DREDI (**Fig. 2**). To quantify how such frustration varies with length scale, we define an in-plane orientation order parameter $O_{hv} = (n_h - n_v)/n$, where $n$ is the total number of domains, and $n_h(n_v)$ denotes the number of horizontal (vertical) domains within a given field of view[48]. The resulting $O_{hv}$ maps capture how in-plane domain anisotropy varies with spatial averaging. Coarse-grained map computed using 10 μm × 10 μm windows show pronounced anisotropy dominated by vertically aligned 71° stripe domains (**Fig. 4A**, background map), whereas finer 1 μm × 1 μm windows reveal localized frustrated regions (white patches in **Fig. 4B**). The corresponding real-space DREDI image (**Fig. 4C**) confirms the coexistence of multiple domain orientations within these mesoscale pockets.

To analyze the spatial organization of these frustrated patches, we performed fractal-dimension analysis on binarized $O_{hv}$ maps, labeling $|O_{hv}| \leq \tau$ (here τ=0.6) as "frustrated". The box-counting results (**Fig. 4D**) reveal two scaling regimes with a clear crossover at ~4 μm. For window sizes <4 μm, the frustrated network follows a fractal dimension $D_F \approx 1.33$ (blue solid line), characteristic of filamentary, self-similar clusters. For windows >4 μm, the dimension approaches $D_F \approx 2$ (red dashed line), indicating an area-filling, effectively 2D percolating mosaic. These results show that the vertex-like polar frustration revealed by DREDI is not a local anomaly but part of a mesoscale-to-macroscale percolating network throughout the film.

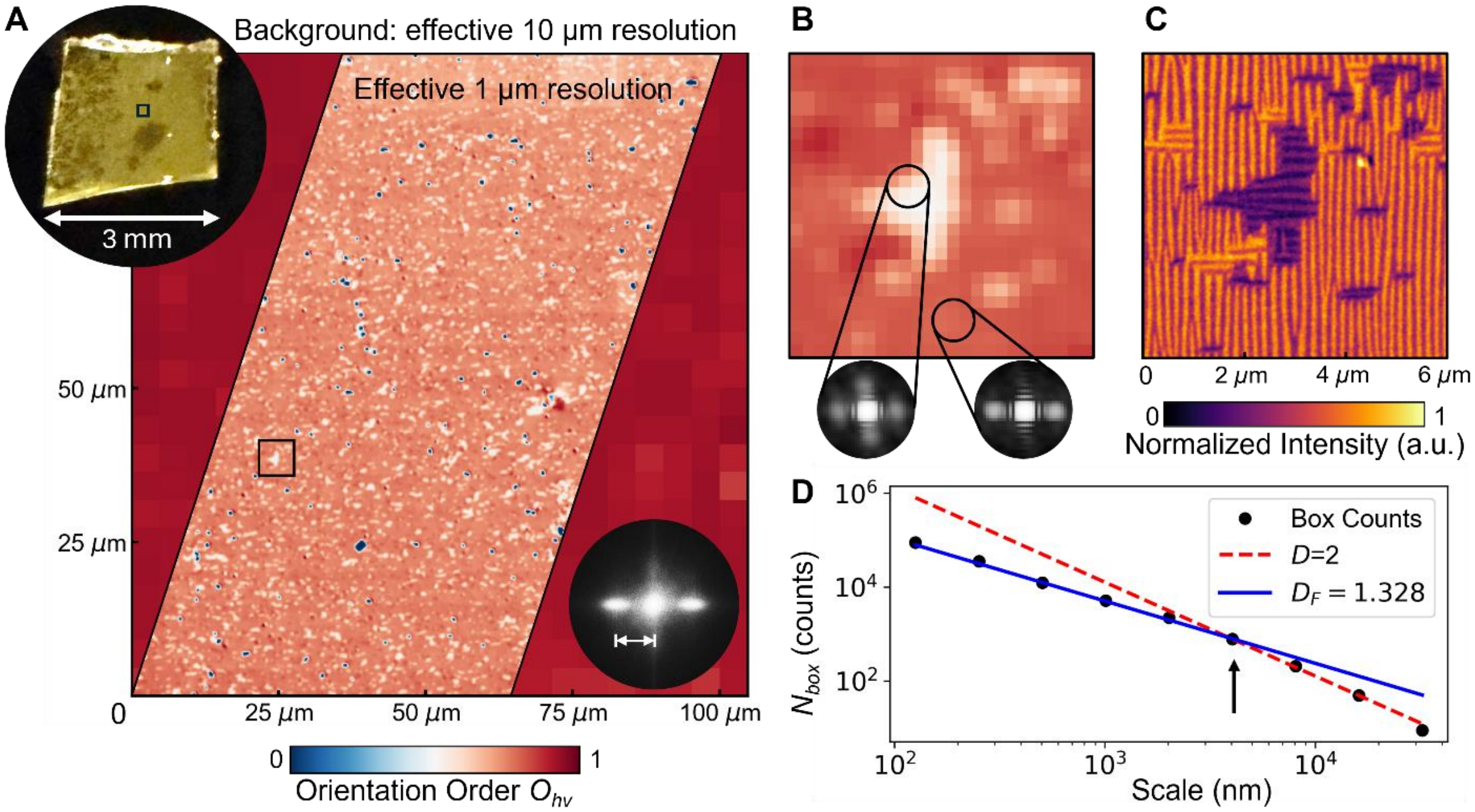

**Fig. 4| Multi-scale mapping of in-plane polarization order. (A)** Spatial averaging dependence of the in-plane orientational order parameter $O_{hv}$ as a function of field-of-view size. Large-area $O_{hv}$ map generated using a 10 μm × 10 μm averaging window, showing global anisotropy dominated by vertical 71° stripes. A swath computed with 1 μm × 1 μm windows reveals localized frustrated regions. Top-left inset: photograph of the ~3 mm $BiFeO_3$ sample, with the black box marking the mapped region. Bottom-right inset: Fourier Transform of the full 120 μm × 122 μm field of view, confirming an average stripe periodicity of 183 nm. **(B)** Zoom-in orientation map showing local frustration (white patches), confirmed by FFTs of sub-regions. **(C)** Real-space DREDI image from the same area demonstrating coexistence of horizontal and vertical stripe orientations. **(D)** Fractal dimension analysis of frustrated regions, showing a transition from filamentary clusters ($D_F \approx 1.33$, blue solid line) below ~4 μm to an area-filling network ($D_F \approx 2$, red dashed line) at larger scales.

## Generality across ferroelectric materials

To assess the broader applicability of DREDI beyond rhombohedral $BiFeO_3$, we applied the technique to two technologically relevant ferroelectric systems with distinct crystal symmetries, polarization magnitudes, coercive fields, and domain-wall types: a thin-film $LiNbO_3$ (TFLN) waveguide device stack[49] and a $BiFeO_3$ thin film coexisting rhombohedral-like and tetragonal-like phases, epitaxially grown on a $LaAlO_3$ substrate[50] (**Supplementary Fig. S7**). These two systems represent contrasting ferroelectric platforms, an integrated photonic device geometry and an oxide film near a morphotropic phase boundary.

In the mixed-phase BFO film, DREDI resolves nanoscale polarization contrast associated with coexisting *R*-like and *T*-like ferroelastic/ferroelectric variants typical of the morphotropic phase boundary regime (**Fig. S7A**), with domain periodicities below 50 nm clearly resolved. In TFLN, DREDI directly visualizes polarization domains written by focused electron-beam exposure within an actual waveguide device stack, producing a box-in-a-box domain configuration (**Fig. S7B**). These results underscore DREDI's potential as a universal, SEM-based metrology platform for ferroics, capable of mapping polarization structures in materials widely used for integrated photonics, piezoelectrics, actuators, and emerging ferroelectric logic nanodevices.

## Discussion

The depth-dependent polarization frustration observed by DREDI can be attributed to the coupled influences of strain relaxation, uncompensated charge, and oxygen octahedral rotations (OOR). Rather than uniform 71˚ stripe domains, both DREDI and MEP reveal embedded, frustrated 3D nanodomains. Cross-sectional four-dimensional (4D) STEM nanodiffraction shows that the $SrRuO_3$ bottom electrode forms ferroelastic twins with distinct OOR variants (**Supplementary Fig. S8**). These twins locally relax epitaxial strain and introduce heterogeneous octahedral connectivity at the $BiFeO_3$/$SrRuO_3$ interface, disrupting the boundary condition for stripe formation. Such sensitivity of ferroelectric domain morphology to interfacial symmetry and strain state is consistent with prior observations in buffer-layer engineered heterostructures[51,52].

We want to note that the DREDI measurements at 15 kV could in principle include contributions from the underlying SRO electrode layer, because higher landing energies increase the electron interaction volume. However, the characteristic length scales of the observed contrast are inconsistent with known structural features of SRO. Cross-sectional 4D-STEM measurements (**Fig. S8**) show that SRO twin domains occur with widths of ~50 nm. In contrast, the frustrated domain features observed in the 15 kV DREDI images (**Fig. 2G**) exhibit lateral dimensions of ~500 nm. This mismatch in characteristic length scales indicates that the dominant diffraction contrast originates from the $BiFeO_3$ layer rather than from the SRO underlayer.

In addition to structural constraints, electrostatic boundary conditions contribute significantly to the observed textures. Many of the domain walls detected by DREDI are inclined, and in some cases exhibit charge accumulation at the boundaries. Phase-field calculations of the in-plane polarization divergence maps, $\vec{\nabla} \cdot \vec{P}$, proportional to the bound charge density, show enhanced bound charge at the cores of vortex and vertex configurations (**Supplementary Fig. S6B-S6D**). These charged or partially compensated domain walls increase the formation energy, making them sensitive to local strain gradients and interfacial screening. The resulting competition between elastic compatibility, octahedral tilt connectivity, and bound-charge minimization provides a natural explanation for the bifurcation of four-fold vortices into asymmetric three-fold vertices near the bottom interface.

Beyond uncovering these buried textures, DREDI provides a broadly accessible platform for quantitative, non-destructive mapping of polar order across multiple length scales. Because it operates on standard SEMs, DREDI enables depth-resolved polarization imaging with acquisition rates orders of magnitude faster than conventional scanning-probe methods, while remaining compatible with *in-line* metrology and *in situ* biasing environments. At present, DREDI employs segmented DBS detectors that capture only the intensity asymmetry in Kikuchi bands. Further integration with emerging reflection Kikuchi diffraction (RKD) geometry[53,54] that record the full momentum distribution of backscattered electrons would extend DREDI to probe quantities like polarization, chirality, symmetry breaking and full 3D lattice deformation fields. This capability would allow *operando* studies of ferroelectric switching, strain-driven phase transitions, and device aging. By bridging the gap between atomic-resolution electron microscopy and wafer-scale inspection, DREDI connects fundamental discovery of topological ferroelectric states with scalable characterization pathways for beyond-CMOS ferroelectric technologies[55,56].

**Acknowledgements**

The authors acknowledge fruitful discussions with Prof. Jian-Min Zuo and Prof. Zijian Hong. We thank Prof. Michelle Povinelli for providing the TFLN waveguide device used in this study. This work was primarily supported by the U.S. Department of Energy, Basic Energy Sciences under award No. DE-SC0025423 (T.-R.L., K.J., A.V.N., Y.-T.S.), with additional support from the USC Viterbi start-up fund. Electron microscopy studies were performed at the Core Center of Excellence in Nano Imaging at the University of Southern California. The authors thank A. Avishai, C. Marks, H. Khant, P. Buenconsejo, and J. Curulli for technical support and careful maintenance of the instruments. S. H., R.R., M.R., and D.G.S. acknowledge that this research was sponsored by the Army Research Laboratory and was accomplished under Cooperative Agreement Number W911NF-24-2-0100. H.S. and J.R. acknowledge support from the USC Annenberg Fellowship. The views and conclusions contained in this document are those of the authors and should not be interpreted as representing the official policies, either expressed or implied, of the Army Research Laboratory or the U.S. Government. The U.S. Government is authorized to reproduce and distribute reprints for Government purposes notwithstanding any copyright notation herein.

## Materials & Methods

### Epitaxial thin-film fabrication

BFO samples (30 nm thick) were deposited on DSO and TSO substrates using both molecular-beam epitaxy (MBE) and pulsed laser deposition (PLD). For PLD-grown samples, a $SrRuO_3$ bottom electrode was included, modifying the electrostatic boundary conditions and stabilizing predominantly 71° domain configurations, in contrast to the 109° domain walls in films without the electrode. MBE films were grown by reactive MBE in a VEECO GEN10 system using a mixture of 80% ozone (distilled) and 20% oxygen. Elemental sources of bismuth and iron were used at fluxes of $1.5\times10^{14}$ and $2\times10^{13}$ atoms per $cm^2s$ respectively. All films were grown at a substrate temperature of 675°C and a chamber pressure of $5\times10^{-6}$ Torr. PLD samples were deposited using a 248 nm KrF laser fluence of 1.8 $Jcm^{-2}$ under a dynamic oxygen pressure of 140 mTorr at 710 °C with a 15 Hz laser pulse repetition rate. The samples were cooled down to room temperature at 30 °C/min at a static $O_2$ atmospheric pressure.

### DREDI imaging

SEM imaging was performed using the Thermo Scientific Helios G5 UX FIB/SEM, equipped with immersion lens mode and a retractable, segmented directional backscatter (DBS) detector. For DBS imaging, the sample is mounted flat, with the detector positioned coaxially with the incident electron beam path along the $[001]_{pc}$ zone axis (i.e., growth direction or plan-view). Along this zone axis, the intensity asymmetry in the Kikuchi bands along the polar axes can be detected by the segmented DBS detectors. DBS images were acquired using a beam current of 0.8 nA and a dwell time of 5 μs. A series of images was collected over a range of accelerating voltages from 2 kV to 25 kV to investigate voltage-dependent contrast behavior. To compare features at the exact same region, images acquired at different landing voltages were registered to sub-pixel accuracy using scale-invariant feature transform (SIFT). Additional details for estimating depths with various electron landing voltages are discussed in *Monte Carlo Simulations on Interaction Volume*.

**STEM and MEP**

The cross-sectional samples were prepared using standard lift-out procedures in a Thermo Fisher Helios G5 Ga FIB-SEM DualBeam system. A standard in situ lift-out approach was used for the cross-sectional sample. To mitigate ion-beam-induced damage, sequential carbon and platinum protective layers were deposited prior to milling. Initial thinning was performed using a 30 kV $Ga^+$ ion beam to reduce the lamella thickness to approximately 100 nm, followed by low energy milling at 5 kV to reach ~50 nm. A final polishing step at 2 kV was conducted to minimize surface amorphization and residual beam damage.

STEM imaging was performed using an aberration-corrected Thermo Fisher Spectra 200 X-CFEG STEM operated at 200 keV, with a probe semi-convergence angle of 25 mrad. Low-angle annular dark field (LAADF) images were acquired with a probe current of 100 pA, collection angle of 15-82 mrad, and dwell time ranging from 100-500 ns per pixel. Thirty frames were collected, registered, and averaged for correcting stage drift.

MEP datasets were acquired using a Dectris ARINA direct electron detector with a frame time of of 35 μs, with probe defocused 25 nm above the sample and scan step size of 0.37 Å, yielding a total dose of ~$1.6\times10^5$ electrons/$Å^2$. The automatic differentiation algorithm in the PtyRAD package was used as the reconstruction algorithm[57]. In the reconstruction, a total number of 20 slices with a slice thickness of 1 nm was used. Ten separate probe modes were used to account for partial coherence of the beam. The presented results were obtained after 2000 iterations with gradual relaxation of the layer regularization along the beam direction. Atomic positions were determined with sub-pixel precision using a combination of center-of-mass refinement and two-dimensional Gaussian fitting, implemented via the Atomap software package[58]. A custom Python script was used to analyze the atomic displacements and extract the polarization vectors of the BFO thin film. The local dipole moments were quantified by measuring the displacement of cations (Bi and Fe atoms) relative to the centroid of the anions (O atoms).

**Phase field simulation**

Two coupled sets of order parameters are introduced in the phase-field modeling: the spontaneous ferroelectric polarization $P_i$ ($i$ = 1-3) and spontaneous oxygen octahedral tilt (OT) $\theta_i$ (with $i$ = 1-3). The temporal evolution of these order parameters in BFO thin film grown on SRO-buffered DSO substrate is described by the time-dependent Ginzburg-Landau equation[59]:

$$\frac{\partial \phi}{\partial t} = -L\frac{\delta F}{\delta \phi}, \#(1)$$

where $\phi$, $t$ and $L$ denote the order parameter (either $P_i$ or $\theta_i$), time step, and kinetic coefficients related to the domain-wall mobility, respectively. The total free energy $F$ accounts for contributions from bulk, gradient, elastic, and electrostatic energies, and is formulated as a volume integral:

$$F = \int \left(f_{bulk}(P_i, \theta_i) + f_{grad}(P_{i,j}, \theta_{i,j}) + f_{elas}(P_i, \theta_i, \varepsilon_{ij}) + f_{elec}(P_i, E_i)\right) dV. \quad (2)$$

The bulk energy density can be written as:

$$f_{bulk} = \alpha_{ij}P_iP_j + \alpha_{ijkl}P_iP_jP_kP_l + \alpha_{ijklmn}P_iP_jP_kP_lP_mP_n + \beta_{ij}\theta_i\theta_j + \beta_{ijkl}\theta_i\theta_j\theta_k\theta_l + t_{ijkl}P_iP_j\theta_k\theta_l, \quad (3)$$

where $\alpha_{ij}$, $\alpha_{ijkl}$, $\alpha_{ijklmn}$, $\beta_{ij}$, and $\beta_{ijkl}$ are the Landau energy coefficients of polarization and OT order parameters, and $t_{ijkl}$ describes their coupling. The gradient energy density can be calculated by:

$$f_{grad} = \frac{1}{2}g_{ijkl}P_{i,j}P_{k,l} + \frac{1}{2}v_{ijkl}\theta_{i,j}\theta_{k,l}, \quad (4)$$

where $g_{ijkl}$ and $v_{ijkl}$ are gradient energy coefficients. The elastic energy density can be expressed as:

$$f_{elas} = \frac{1}{2}c_{ijkl}(\varepsilon_{ij} - \varepsilon_{ij}^0)(\varepsilon_{kl} - \varepsilon_{kl}^0), \quad (5)$$

where $c_{ijkl}$ is the elastic stiffness tensor, $\varepsilon_{ij}$ is the total strain. The eigenstrain $\varepsilon_{ij}^0$ is related to the polarization and OT order parameters by the coupling coefficients $Q_{ijkl}$ and $\Lambda_{ijkl}$ via $\varepsilon_{ij}^0 = Q_{ijkl}P_kP_l + \Lambda_{ijkl}\theta_k\theta_l$.

The electrostatic energy density is given by:

$$f_{elec} = -\frac{1}{2}\kappa_0\kappa_bE_iE_j - E_iP_i, \quad (6)$$

where $\kappa_0$ and $\kappa_b$ are the vacuum and relative permittivities, respectively, and $E_i$ is the total electric field.

A discrete three-dimensional mesh of $100\Delta x \times 100\Delta y \times 110\Delta z$ grid points was used, with a real-space grid spacing of $\Delta x = \Delta y = 2.0$ nm and $\Delta z = 1.0$ nm. The model system consists of 30 grids for the substrate layer, 50 grids for the BFO thin film, and 30 grids for the air layer, stacked sequentially from bottom to top. Periodic boundary conditions were imposed in the in-plane dimensions, while a superposition method was implemented along the thickness dimension[60]. The mechanical boundary condition was defined such that the out-of-plane stress is fully released on the top of the film, while the out-of-plane displacement is zero at the bottom of the substrate sufficiently far away from the substrate/film interface. The pseudocubic lattice parameter for BFO film was taken as 3.965 Å, while those of $(001)_{pc}$-DSO substrate is $a = 3.952$ Å, $b = c = 3.947$ Å. A short-circuit electrical boundary condition was applied. The initial polarization state was generated by introducing small-amplitude random noise to mimic thermal fluctuations, combined with a weak net *x*-directional perturbation at the top region of the BFO film to promote the formation of experimentally observed ordered 71° domains. The material parameters for BFO are available in the literatures[61] and listed in **Supplementary Table S1**. The simulations were performed at a temperature of 298 K with a normalized time step of 0.01.

**Supplementary Information for**

**Revealing buried ferroelectric topologies by depth-resolved electron diffraction imaging**

Ting-Ran Liu[1], Koushik Jagadish[1], Xiangwei Guo[2,3], Maya Ramesh[4], Peter Meisenheimer[5], Harish Kumarasubramanian[1], Sajid Husain[5], Ann V. Ngo[1], Amir Avishai[6], Jayakanth Ravichandran[1,6,7], Darrell G. Schlom[4], Ramamoorthy Ramesh[5], Yu-Tsun Shao[1,6]*

Corresponding author: Yu-Tsun Shao
Email: yutsunsh@usc.edu

**This Word file includes:**

Supplementary text
Figures S1 to S12
Table S1
SI References

**Supplementary Information Text**

**Supplementary Note 1. Polarization direction determination.** To have a more careful analysis of the diffraction contrast in DREDI, we (1) performed dynamical diffraction simulations of backscattered electrons to identify polarity-sensitive Kikuchi bands, and (2) calibrated the collection angles of the DBS detector segments in correspondence with the simulated diffraction patterns. The dynamical diffraction simulations were carried out using the Bloch-wave formalism implemented in the EMsoft package[1]. Electron channeling patterns (ECP) were simulated for $BiFeO_3$ crystals with space group R3c and at 0° stage tilt, allowing us to identify polarity-sensitive Kikuchi bands **(Fig. S1)**.

In addition, we calibrated the angular collection ranges of the DBS detector segments **(Fig. S3)**. The inner and outer angles were estimated to be approximately 22.8° and 38.7°, respectively, based on the measured detector geometry and effective sample-to-detector distance using arctan(detector-radius / detector-to-sample distance).

In the present work, both DREDI experiments and Bloch-wave ECP simulations were performed under the same zero-sample-tilt geometry, with the detector positioned around the pole piece. Under this configuration, the backscattered electrons detected at large scattering angles (157.2°-141.3° for the DBS detector) remain close to the incident beam energy. To verify this, we performed additional ECP simulations at 2, 5, 10, and 15 kV, which show no contrast inversion **(Fig. S1 & S9)**. This is consistent with prior reports that the ECP patterns are largely contributed from backscattered electrons within ~1 kV of the incident energy[2,3].

**Supplementary Note 2. Comparison between ECCI and DREDI.** In electron channeling contrast imaging (ECCI), the contrast primarily arises from orientation-dependent electron channeling, often enhanced by a controlled mistilt angle to reveal crystalline defects like dislocations or ferroelastic domains. On the other hand, DREDI captures the breaking of Freidel symmetry in non-centrosymmetric crystals due to dynamical diffraction effects, where intensity asymmetry between Kikuchi bands ($\pm\vec{G}$) can be used to determine polarization direction.

To distinguish contributions from crystal orientation and polarity, we performed sample rotation experiments **(Fig. S10)**. Here, the relative azimuthal angles between the sample substrate crystallographic planes and the detector is systematically varied while maintaining the same detector collection angle. This rotation shifts the azimuthal angle of the diffraction asymmetry across detector segments, leading to corresponding changes in domain contrast. Such behavior is consistent with polarity-sensitive asymmetry in Kikuchi bands, rather than solely orientation-dependent channeling contrast.

In addition, Bloch-wave simulations of electron channeling patterns **(Figs. S1 & S9)** identify specific Kikuchi bands that exhibit polarity-sensitive intensity asymmetry, further supporting this interpretation.

**Supplementary Note 3. Depth-sensitivity of DREDI.** The depth sensitivity of DREDI is governed by the electron interaction volume and the depth-dependent distribution of backscattered electrons. A first-order estimate of the electron penetration depth can be obtained using the semi-empirical Kanaya-Okayama (K-O) range formula, with expression given by:

$$R = 0.0276\,[AE_0^{1.67}]/[\rho Z^{0.89}]$$

where R is the penetration depth (μm), $E_0$ is the beam energy (keV), and A, Z, and ρ are the atomic weight, atomic number, and density (g/cm$^3$) of the material, respectively. This provides an estimate of the overall interaction volume.

However, since DREDI detects the backscattered electrons, the effective depth sensitivity is more directly related to the depth distribution of backscattered electrons rather than the total interaction range. This depth dependence is not captured analytically by the K-O expression but can be estimated using Monte Carlo simulations (e.g. CASINO).

As shown in **Fig. S5**, the backscattered electron distribution shifts to greater depths and broadens with increasing landing energy. The effective depth sensitivity of DREDI can therefore be interpreted as the depth interval over which the backscattered electron distribution contributes significantly to the detected signal.

**Monte Carlo simulations on interaction volume.** Monte Carlo simulations were performed using the CASINO software package[4] to model the spatial distribution of electron energy deposition within the sample. The simulated sample was constructed to reflect the experimental 30 nm $BiFeO_3$/ 30 nm $SrRuO_3$/$DyScO_3$ substrate heterostructure, matching both the material composition and the individual layer thicknesses. Simulations were conducted with incident electron beam energies of 2, 5, 10, and 15 keV, using a beam diameter of 1 nm. For each condition, 2,000,000 electrons were simulated to ensure statistical robustness in the calculated energy density profiles.

**Electron Backscattered Diffraction (EBSD).** EBSD data were acquired at a working distance of 4 mm, with the sample mounted on a standard 70° pre-tilted stub to enhance pattern acquisition efficiency. No additional surface preparation was performed, in order to preserve the atomically smooth morphology of the as-grown films. Measurements were conducted using a Thermo Scientific Helios G4 UXe PFIB/SEM equipped with an Oxford Symmetry S2 EBSD setup. EBSD were performed with an accelerating voltage of 5–10 kV, a beam current of 2 nA, and a step size of 10-50 nm over a 10 μm × 10 μm area. Dynamical diffraction simulations of Kikuchi patterns were carried out using AztecCrystal software using the Bloch-wave algorithm. Post-processing of the 4D EBSD datasets—comprising full Kikuchi diffraction patterns ($k_x$, $k_y$) acquired at each scan position (*x, y*)—was performed using principal component analysis (PCA) and k-means clustering. The first 30 components were selected from the PCA analysis to reduce noise in the raw patterns. The subsequent denoised data were used as input for k-means clustering to segment different regions in the BFO thin film into distinct crystallographic domains. The number of clusters was determined empirically by evaluating the pattern variance across different regions and selecting the value that yielded physically meaningful domain boundaries. Both PCA and clustering were implemented using custom Python scripts.

**3D rendering methodology for unit cell tetragonality from MEP.** The 3D rendering in Figure 2J is constructed from depth-resolved multislice electron ptychography (MEP) data. MEP reconstructs the projected potential slice-by-slice along the electron beam direction, allowing us to resolve the atomic structure at different depths.

For each reconstructed slice, we extract a tetragonality map by measuring the distances ratio between Bi atomic columns for each unit cell. This results in a stack 20 depth-resolved tetragonality maps along the $[010]_{pc}$ direction. The 3D rendering plot was processed with the "z-project" function in ImageJ software.

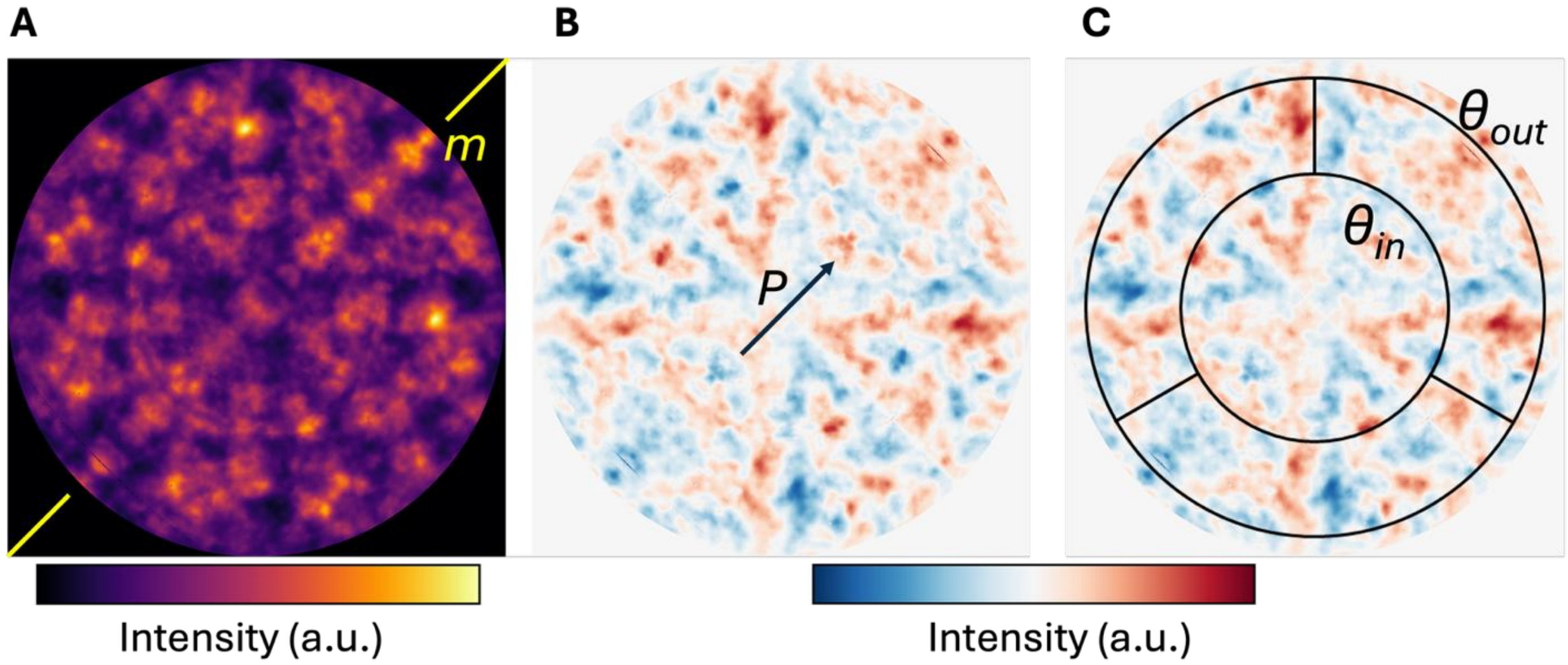

**Fig. S1| Simulated electron channeling patterns and detector geometry used for DREDI analysis.** (**A**) Bloch-wave simulated electron channeling pattern (ECP) of $BiFeO_3$ viewed along the pseudocubic [001] direction at a landing energy of 2 kV. *m* denotes the mirror plane. (**B**) Intensity difference perpendicular to the mirror plane highlighting polarity-sensitive diffraction contrast along the polar axis. The arrow indicates the polarization direction *P* parallel to pseudocubic [110]. (**C**) Overlay of the ECP intensity difference map with the segmented DBS detector geometry used in the experiment. $\theta_{in}$ and $\theta_{out}$ denote the inner and outer collection angles of the DBS detector, respectively.

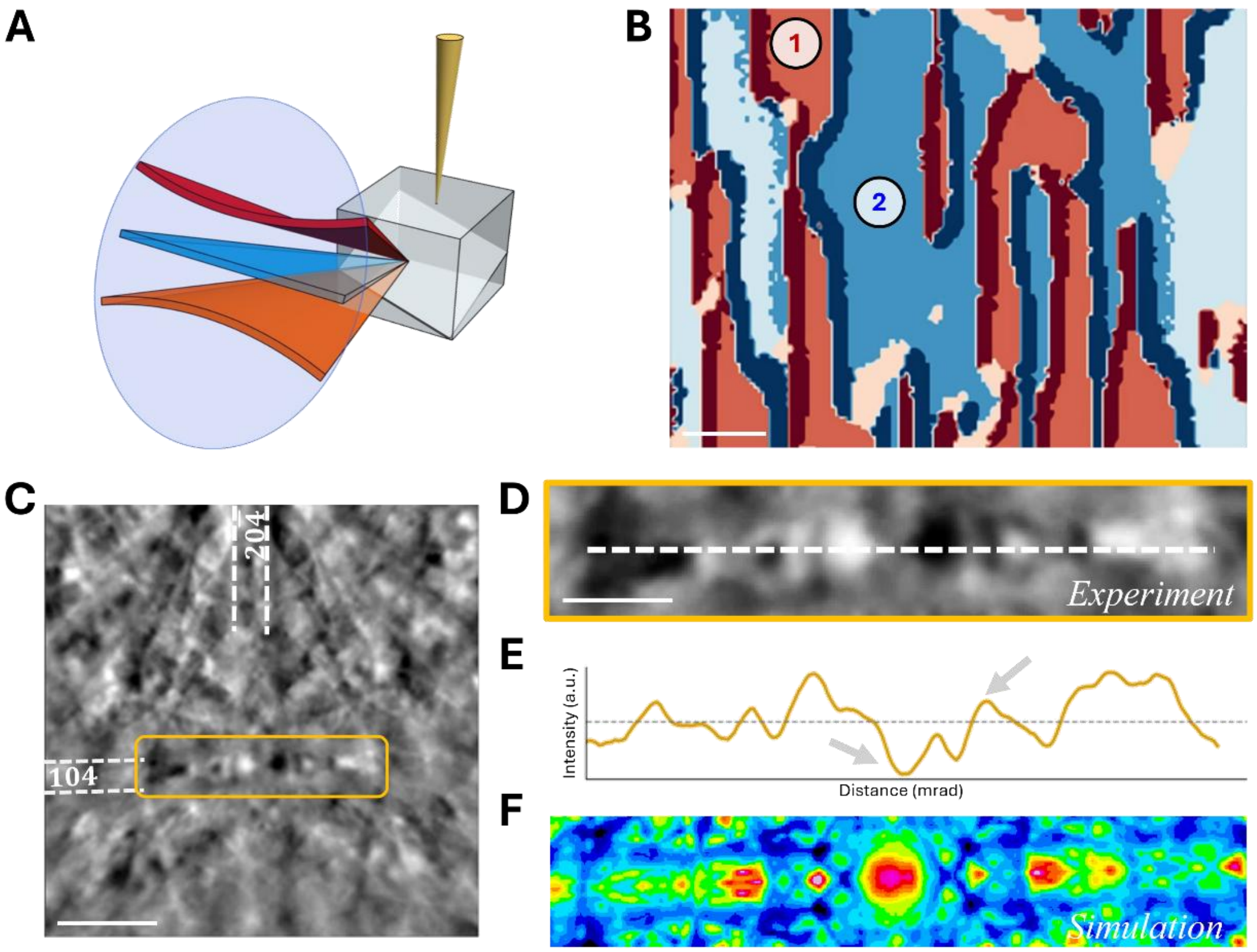

**Fig. S2| Polarization imaging from Friedel symmetry breaking using EBSD. (A)** Schematic of the EBSD geometry. Four-dimensional datasets (*x, y, $k_x$, $k_y$*) were acquired using a CMOS detector. **(B)** Ferroelectric 109° domain configuration in $BiFeO_3$ can be reconstructed from the EBSD datasets. **(C)** Difference map of representative Kikuchi patterns from domains 1 and 2, as labelled in (B). **(D)** Enlarged area of the difference patterns, from yellow box in (C). **(E)** Corresponding line profile of (D). Intensity asymmetries in the Kikuchi bands highlighted with arrows indicate direct evidence of Friedel symmetry breaking in non-centrosymmetric $BiFeO_3$. **(F)** Dynamical Bloch-wave simulation reproduces similar Kikuchi band asymmetries, supporting the experimental observations. Scale bars: 50 mrad (C), 20 mrad (D).

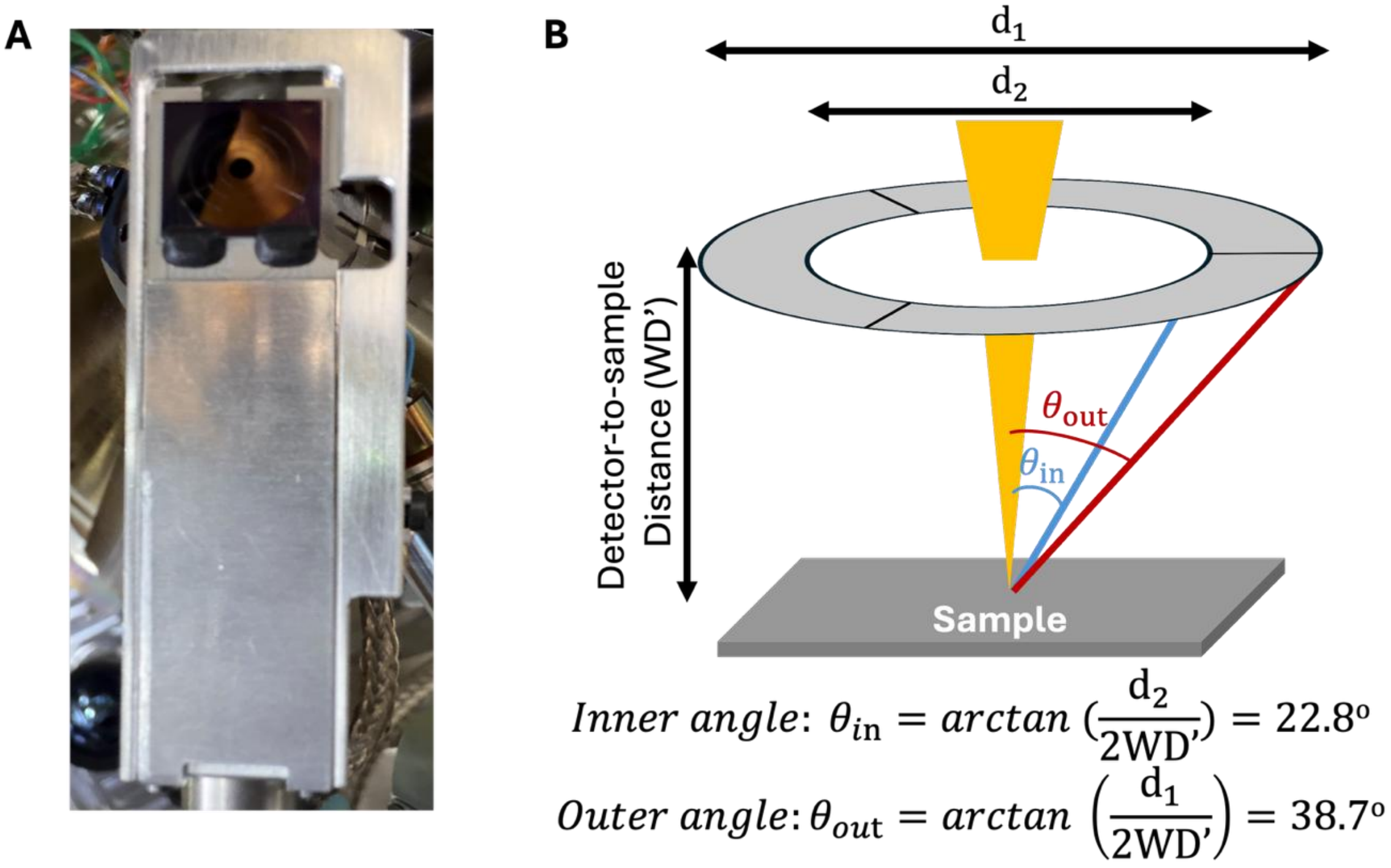

**Fig. S3| Collection angle of the DBS detector in DREDI.** **(A)**Photo of the DBS detector in the instrument. **(B)** Schematic and calculation of the collection angle. The collection angle is 22.8º ~ 38.7º.

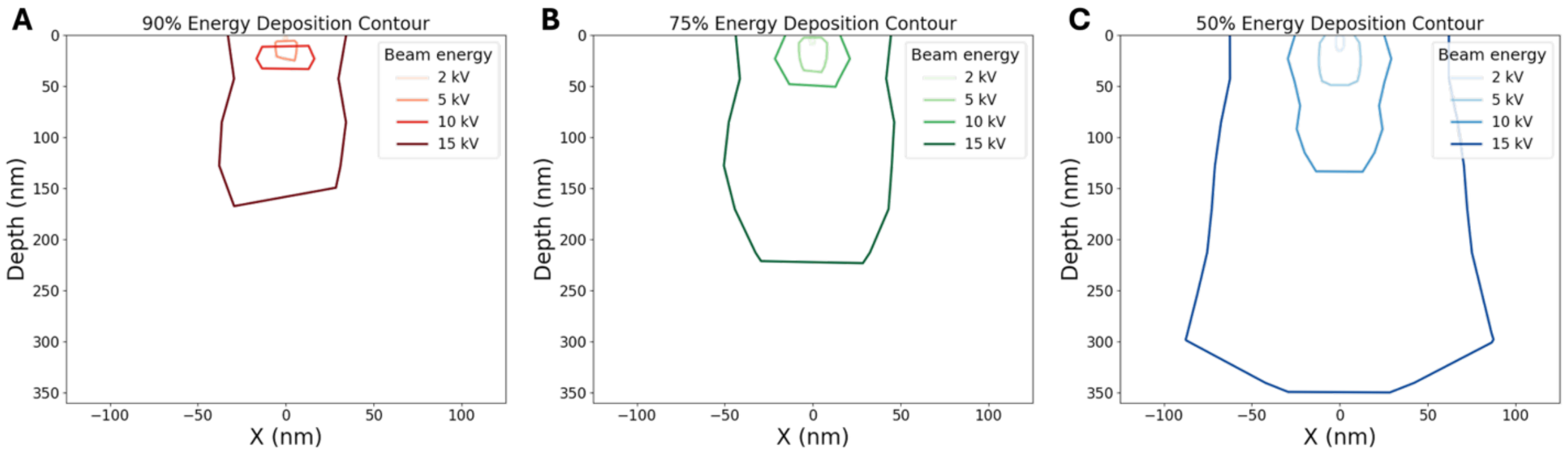

**Fig. S4| Monte-Carlo simulation of electron energy distribution.** Simulated energy distributions of incident electron beams at various accelerating voltages using CASINO. The simulated structure is a 30 nm-$BiFeO_3$/ 30 nm-$SrRuO_3$/ $DyScO_3$ heterostructure. **(A-C)** Corresponds to the energy deposition of 90%, 75%, and 50% of initial electron energy. Four different accelerating voltages, 2 kV, 5 kV, 10 kV, and 15 kV were simulated.

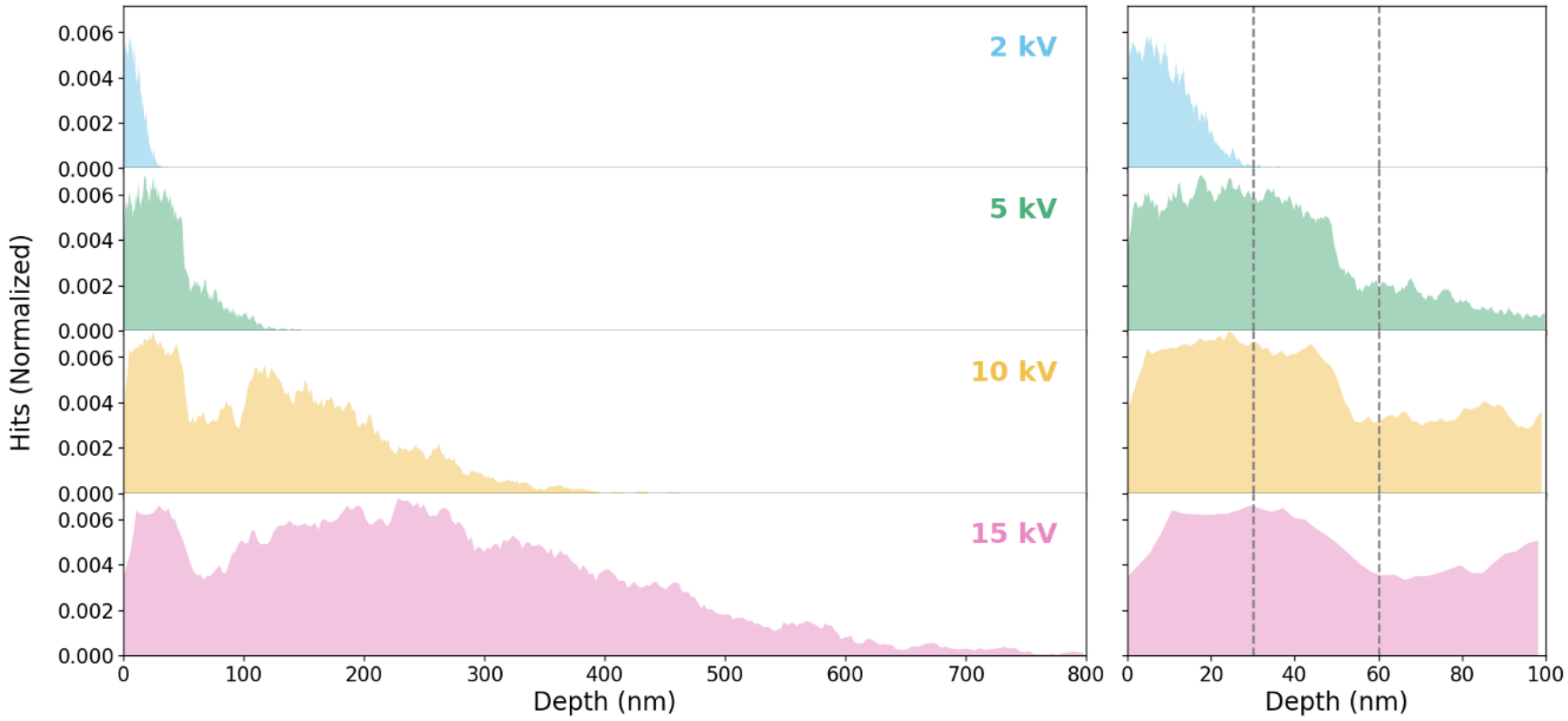

**Fig. S5| Monte-Carlo simulation of backscattered electron distribution.** Simulated backscattered electron penetration depth at various accelerating voltages (2 kV, 5 kV, 10 kV, and 15 kV) using CASINO. **(A)** Full range of penetration depth, **(B)** enlarged range from **(A)**. Marked gray lines in **(B)** correspond to the experimental heterostructure interfaces, $BiFeO_3/SrRuO_3$ (30 nm) and $SrRuO_3/DyScO_3$ (60 nm) in the DREDI setup.

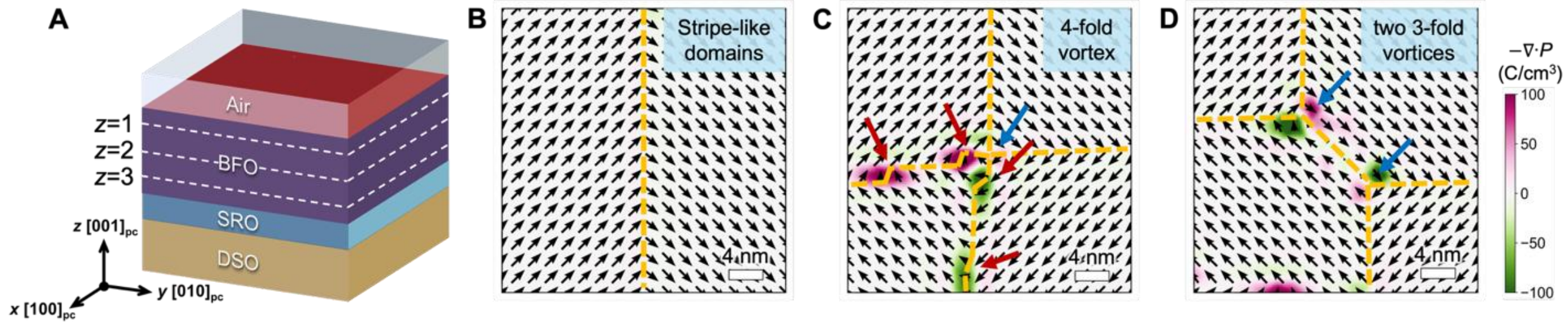

**Fig. S6| Bound charge density analysis of domain walls from phase-field simulation.** Schematic illustration of the simulated heterostructure consisting of BFO/SRO/DSO layers with air on top. **(B)** Near the top surface (z=1 nm), the straight, stripe-like 71° domain walls show nearly no bound charges. **(C)** In the middle (z=19 nm), the 4-fold vortex core with some stepped 71° domain walls exhibit strong bound charges, confirming the high-energy metastability of this topology. **(D)** Near the bottom interface (z=30 nm), domain splitting into two 3-fold vortices releases the bound charges at the core, while the 71° domain walls recover their straight and stable configuration.

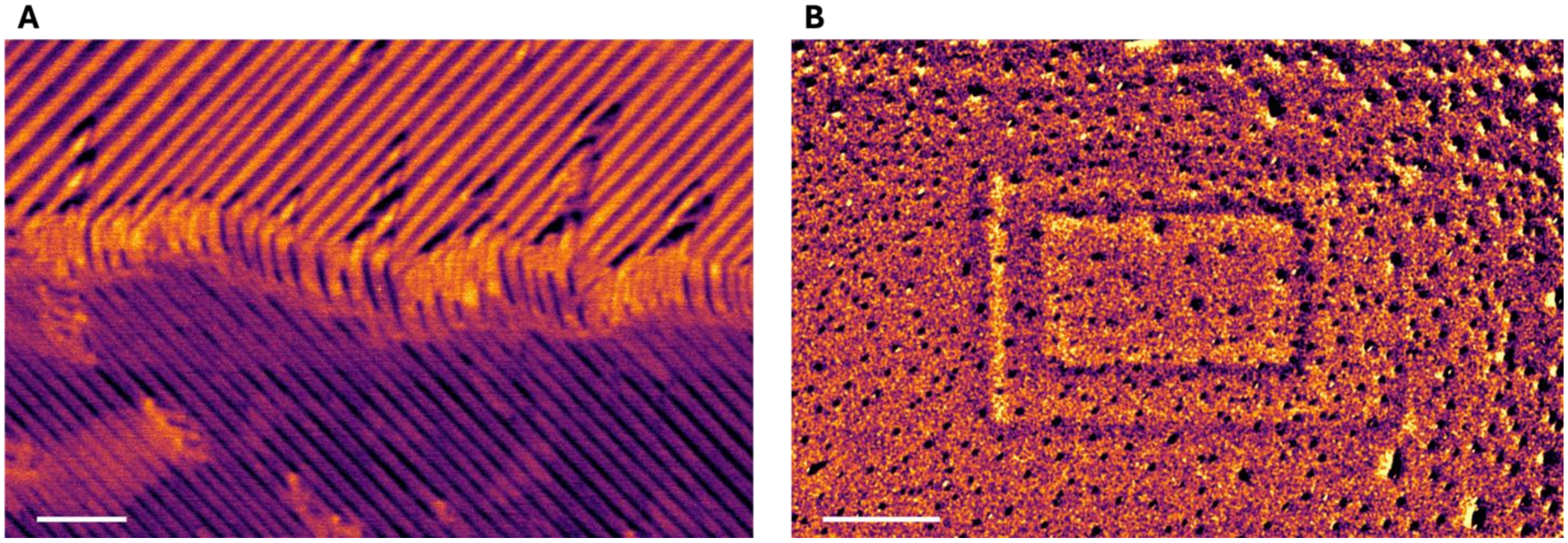

**Fig. S7| Generality of DREDI across ferroelectric materials and device platforms. (A)** DREDI mapping of nanoscale morphotropic phase boundary (MPB) in tetragonal $BiFeO_3$ (*T*-BFO), resolving ferroelastic/ferroelectric domain variants. Domains with periodicities of <50 nm (top, $M_C$/*T*-like phase) and <90 nm (bottom, $M_A$/*R*-like phase) can be clearly resolved. The middle stripe represents the MPB region where both phases coexist. **(B)** DREDI imaging of polarization switching in a thin-film $LiNbO_3$ (TFLN) waveguide device stack, revealing a box-in-a-box domain configuration written by a focused electron beam. Dark dots are intrinsic defects in the film. These examples demonstrate the applicability of DREDI to ferroelectric materials with distinct crystal symmetries and to functional device architectures. Images are used to illustrate domain structure and morphology. Scale bars: 500 nm (A), 5 um (B).

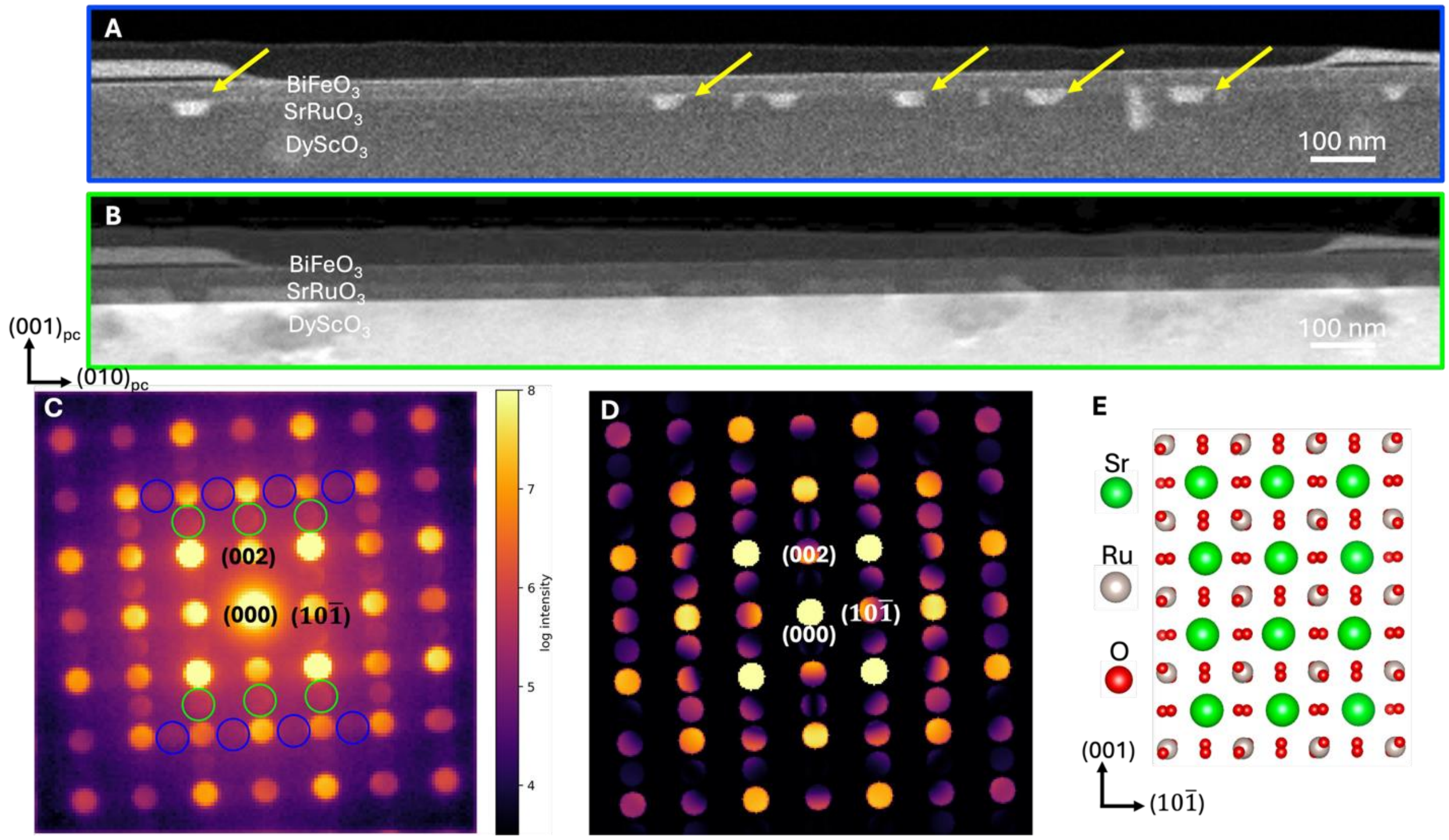

**Fig. S8| $SrRuO_3$ ferroelastic twin domains revealed by scanning electron nanodiffraction.** (A, B) Virtual dark-field image constructed from Bragg reflections related to oxygen octahedral rotation (OOR) patterns in $SrRuO_3$ structure. The bright regions in the $SrRuO_3$ layer (yellow arrows) indicate the presence of ferroelastic twins. (C) Mean diffraction pattern of the $SrRuO_3$ layer showing OOR-sensitive Bragg reflections predominantly parallel to (002), with some weak reflections along the perpendicular orientation. The blue and green circles indicate the Bragg reflections used for generating dark-field image in (A) and (B). (D) Bloch-wave simulation of $SrRuO_3$ along the [101] zone axis, with thickness of 20 nm. (E) Structure model of $SrRuO_3$ with Pnma symmetry viewed along the [101] zone axis, exhibiting OOR that contributes to the doubling of perovskite unit cell.

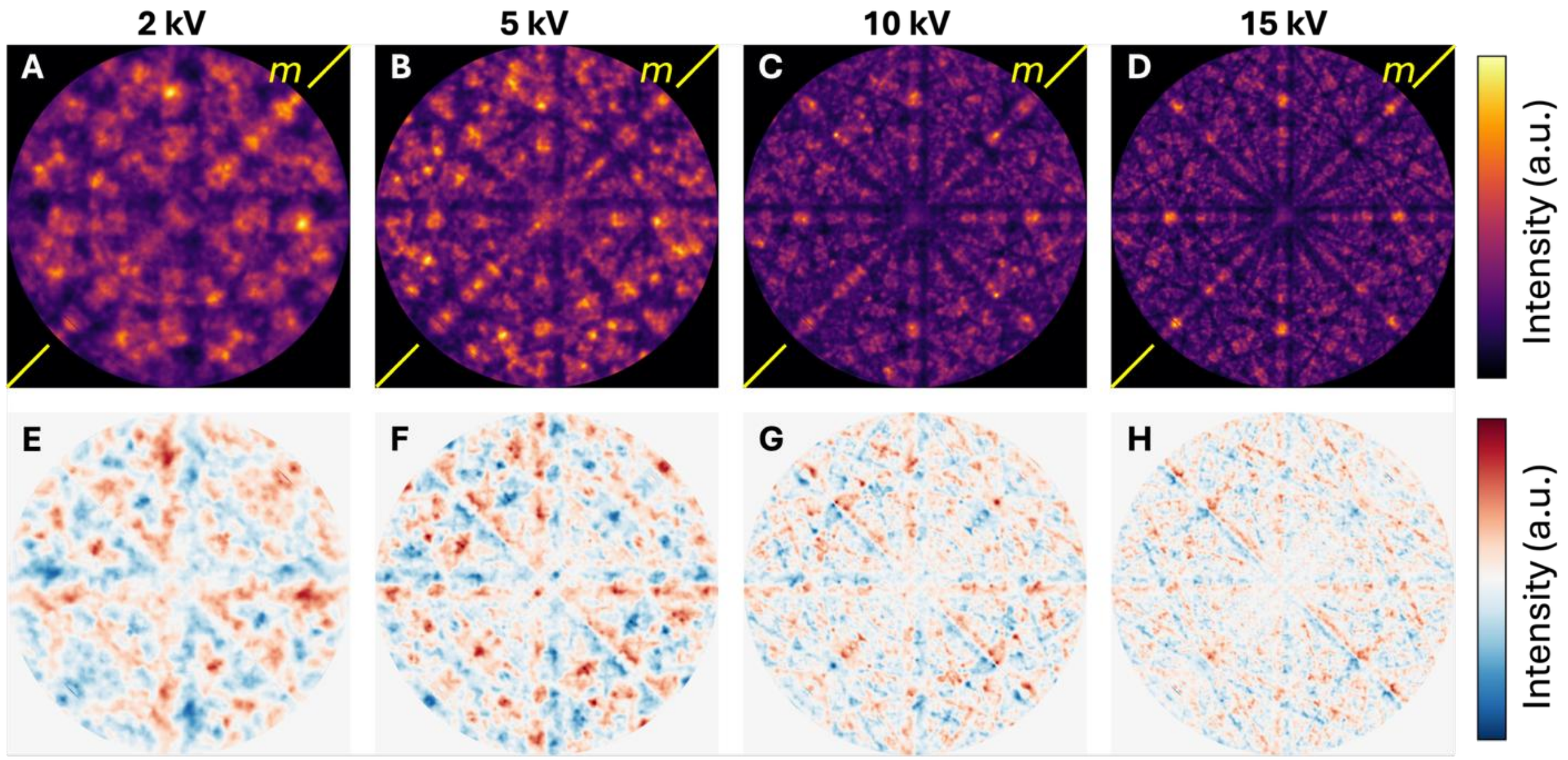

**Fig. S9| Polarity direction in electron channeling patterns at different landing energies.** (**A-D**) Bloch-wave simulated electron channeling pattern (ECP) of $BiFeO_3$ viewed along the pseudocubic [001] direction at landing energies of 2 kV, 5 kV, 10 kV, and 15 kV, respectively. The yellow line labeled *m* denotes the mirror plane. (**E-H**) Corresponding intensity difference maps calculated perpendicular to the mirror plane, highlighting polarity-sensitive diffraction contrast along the polar axis. In all cases, the center of mass of the ECP consistently points toward the pseudocubic [110] direction (top right), corresponding to the polarization direction *P*.

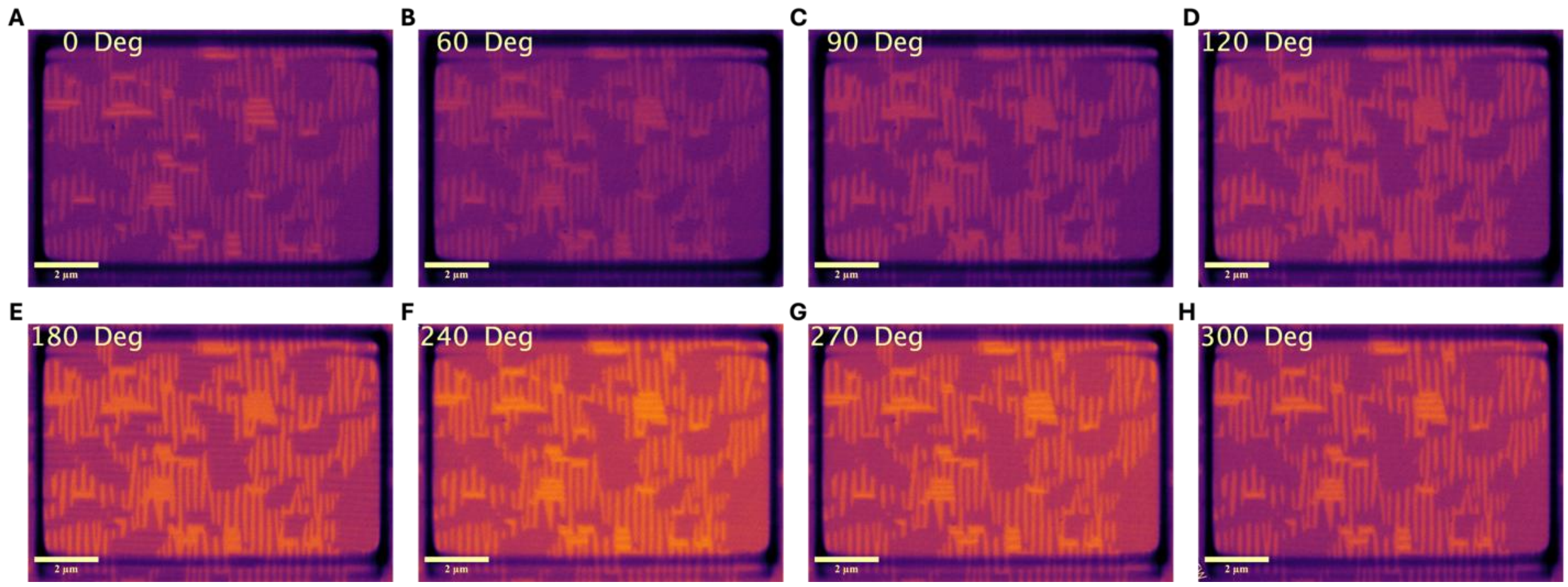

**Fig. S10| Variant domain contrast from sample-detector relative orientations.** DBS detector collects the intensity asymmetries between Friedel pairs of Kikuchi bands ($\pm\vec{G}$). The relative orientation of the domain to the detector would affect signal collection in DBS and thus result in different domain contrast. (A-H) shows domain contrast from the identical region of interest under varying relative orientation between the sample and the DBS detector. Stage rotation was used to vary the sample-detector orientations.

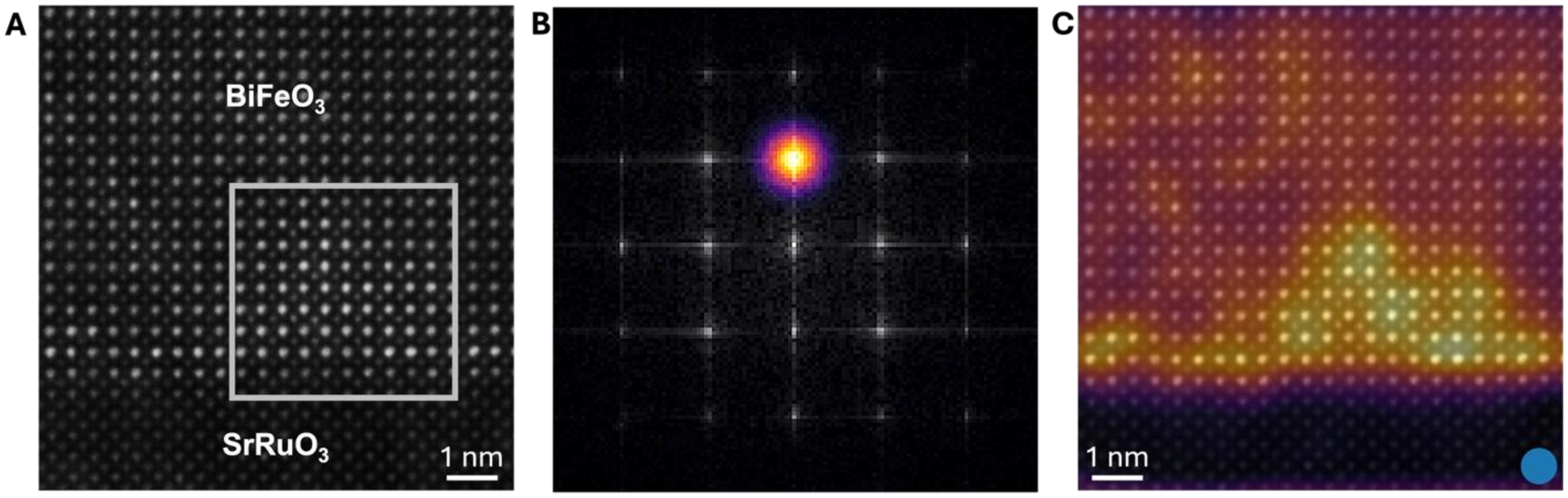

**Fig. S11| Fourier filtering of vertex-like polar texture. (A)** LAADF-STEM image from Figure 3 in main text. **(B)** Power spectrum of (A) and the masked signal for Fourier filtering. **(C)** Amplitude of Fourier filtered signal of (A). The Fourier filtering was conducted using the phase lock-in method, with regions indicated in (B). The region of interest (ROI, white box in A) has stronger diffraction contrast correspond to larger lattice distortions in Main Figure 3C.

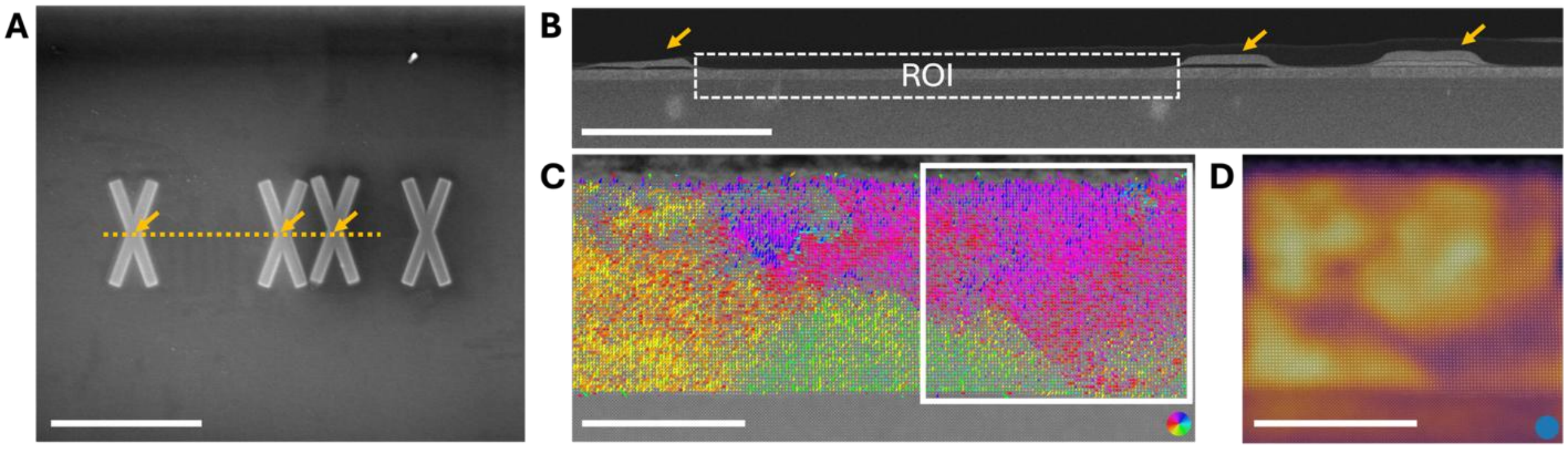

**Fig. S12| Region of interest for cross-sectional STEM.** The region of interest for embedded 3D domain configurations was identified using voltage-dependent DBS images from plan-view, then performed cross-sectional FIB and STEM imaging. (**A**) SEM image of the same region in Fig. 2, where fiducial markers (indicated by yellow arrows) were created by ion deposition prior to destructive FIB milling process. The FIB cut was made along the $[100]_{pc}$ (dashed line in (A)). (**B**) Low magnification cross-sectional STEM image showing the corresponding fiducial markers (yellow arrows) marked by FIB. **(C)** LAADF-STEM image acquired within the region of interest (ROI), showing diffraction contrast sensitive to structural distortions. Overlayed polarization maps shows different polarization orientations, with domain configurations varying along the thin film growth axis. (**D**) Amplitude of Fourier filtered signal from cropped region in (C). Scale bars: 2 μm (A), 1 μm (B), 20 nm (C, D).

**Table S1. Material parameters of BFO film used in the phase-field simulations.**

| $\boldsymbol{\alpha_{11}}$ | **$-3.580 \times 10^{8}\ \mathrm{C^{-2}\cdot m^{2}\cdot N}$** | $\boldsymbol{v_{1111}}$ | **$7.840 \times 10^{-11}\ \mathrm{rad^{-2}\cdot N}$** |
|---|---|---|---|
| $\boldsymbol{\alpha_{1111}}$ | $3.000 \times 10^{8}\ \mathrm{C^{-4}\cdot m^{6}\cdot N}$ | $v_{1122}$ | $-5.138 \times 10^{-9}\ \mathrm{rad^{-2}\cdot N}$ |
| $\boldsymbol{\alpha_{1122}}$ | $1.188 \times 10^{8}\ \mathrm{C^{-4}\cdot m^{6}\cdot N}$ | $v_{1212}$ | $4.977 \times 10^{-9}\ \mathrm{rad^{-2}\cdot N}$ |
| $\boldsymbol{\beta_{11}}$ | $-5.400 \times 10^{9}\ \mathrm{rad^{-2}\cdot m^{-2}\cdot N}$ | $c_{1111}$ | $2.280 \times 10^{11}\ \mathrm{N\cdot m^{-2}}$ |
| $\boldsymbol{\beta_{1111}}$ | $3.440 \times 10^{10}\ \mathrm{rad^{-4}\cdot m^{-2}\cdot N}$ | $c_{1122}$ | $1.280 \times 10^{11}\ \mathrm{N\cdot m^{-2}}$ |
| $\boldsymbol{\beta_{1122}}$ | $6.799 \times 10^{10}\ \mathrm{rad^{-4}\cdot m^{-2}\cdot N}$ | $c_{1212}$ | $0.650 \times 10^{11}\ \mathrm{N\cdot m^{-2}}$ |
| $\boldsymbol{t_{1111}}$ | $4.532 \times 10^{9}\ \mathrm{C^{-2}\cdot rad^{-2}\cdot m^{2}\cdot N}$ | $\Lambda_{1111}$ | $0.08416\ \mathrm{rad^{-2}}$ |
| $\boldsymbol{t_{1122}}$ | $2.266 \times 10^{9}\ \mathrm{C^{-2}\cdot rad^{-2}\cdot m^{2}\cdot N}$ | $\Lambda_{1122}$ | $-0.09200\ \mathrm{rad^{-2}}$ |
| $\boldsymbol{t_{1212}}$ | $-4.840 \times 10^{9}\ \mathrm{C^{-2}\cdot rad^{-2}\cdot m^{2}\cdot N}$ | $\Lambda_{1212}$ | $0.3192\ \mathrm{rad^{-2}}$ |
| $\boldsymbol{g_{1111}}$ | $4.335 \times 10^{-11}\ \mathrm{C^{-2}\cdot m^{4}\cdot N}$ | $Q_{1111}$ | $0.05700\ \mathrm{C^{-2}\cdot m^{4}}$ |
| $\boldsymbol{g_{1122}}$ | $-3.400 \times 10^{-12}\ \mathrm{C^{-2}\cdot m^{4}\cdot N}$ | $Q_{1122}$ | $-0.02000\ \mathrm{C^{-2}\cdot m^{4}}$ |
| $\boldsymbol{g_{1212}}$ | $3.400 \times 10^{-12}\ \mathrm{C^{-2}\cdot m^{4}\cdot N}$ | $Q_{1212}$ | $-0.0007300\ \mathrm{C^{-2}\cdot m^{4}}$ |

**SI References**